\documentstyle[12pt]{article}
\begin{document}

\title{The generation of an entangled four-photon state from two
pairs of entangled two-photon states by using linear optical
elements}
\author{XuBo Zou, K. Pahlke and W. Mathis  \\
\\Institute TET, University of Hannover,\\
Appelstr. 9A, 30167 Hannover, Germany }
\date{}

\maketitle

\begin{abstract}
{\normalsize We present a scheme to produce an entangled
four-photon state from two pairs of entangled two-photon states.
Such entangled four-photon states are equivalent to the quantum
state of two maximally entangled spin-1 particles. The scheme can
also be generated to prepare an entangled $2N$-photon state from
$N$ pairs of entangled two-photon states. Such multi-photon states
play a crucial role in fundamental tests of quantum mechanics
versus local realism and in many quantum information and quantum
computation schemes.

 }

\end{abstract}
The generation of entangled states plays a central role in quantum
optics. An experimental realization in this context can be
achieved with trapped ions \cite{qkc}, cavity QED \cite{ar} and
Bose-Einstein condenses \cite{am}. Many schemes were proposed for
the purpose of generating entangled quantum states atoms
 \cite{sch}. In fact, the GHZ-state was demonstrated with the
trapped ion stored in an atom cavity system \cite{ss,ar}. A broad
field of research is opened by experiments performed with
entangled photons. Photon entanglement can be used to test Bell's
inequality \cite{daa} or to implement quantum information
protocols for quantum teleportation \cite{db}, quantum dense
coding \cite{km} and quantum cryptography \cite{ds}. Recently, an
experimental realization of GHZ-states by means of three or four
photons was reported \cite{dj}. The efficient quantum computation
with linear optics has been put forward \cite{erg}. This kind of
scheme can be used directly to generate photon polarization
entanglement. A sequence of beam splitters is arranged carefully
in order to implement a basic non-deterministic gate. Another
feasible linear optical scheme \cite{zjy} was proposed to produce
photon polarization entanglement with the help of single-photon
quantum non-demolition measurements on an atom-cavity system
\cite{pjj}. In a recent paper \cite{zxb} we proposed a scheme to
generate entangled $N$-photon states of the form
$\frac{1}{\sqrt{2}}(|0,N\rangle+|N,0\rangle)$ via linear optical
elements. Recently, the incentive to produce entangled states of
spin-s objects $(S>1/2)$ has been increased significantly by the
advance of the fields of quantum communication and quantum
information processing \cite{gp,kp,ddd}. Stimulated by
experiment\cite{dj}, in this paper, we present a scheme to
generate an entangled four-photon state from two pairs of
entangled two-photon states
\begin{eqnarray}
\Psi=\frac{1}{\sqrt{3}}(|2H\rangle_1|2V\rangle_2+|H\rangle_1|V\rangle_1
|H\rangle_2|V\rangle_2+|2V\rangle_1|2H\rangle_2) \label{1}
\end{eqnarray}
by using linear optical elements. This entangled four-photon state
is equivalent to two maximally entangled spin-1 particles.
Drummond \cite{pdd} demonstrated the violation of the Bell
inequality by this kind of multi-photon states. The experimental
setup is shown schematically in Fig.1. We assume that two pairs of
maximally entangled two-photon states has been prepared:
$\frac{1}{\sqrt{2}}(|H\rangle_1|V\rangle_2+|V\rangle_1|H\rangle_2)$
and
$\frac{1}{\sqrt{2}}(|H\rangle_3|V\rangle_4+|V\rangle_3|H\rangle_4)$.
The initial state of the system is
\begin{eqnarray}
\Psi_0=\frac{1}{2}(|H\rangle_1|V\rangle_2+|V\rangle_1|H\rangle_2)(|H\rangle_3|V\rangle_4+|V\rangle_3|H\rangle_4)
\,.\label{2}
\end{eqnarray}
Let the modes $H_1$, $H_3$, $V_1$, $V_3$, $H_2$, $H_4$, $V_2$ and
$V_4$ pass through four beam splitters $BS$ of the same kind. The
state of the system evolves into
\begin{eqnarray}
\Psi_1&=&\frac{1}{8}[(\cos\theta
a_{H_1^{\prime}}^{\dagger}+\sin\theta
a_{H_3^{\prime}}^{\dagger})(\cos\theta
a_{V_2^{\prime}}^{\dagger}+\sin\theta
a_{V_4^{\prime}}^{\dagger})\nonumber\\
&& + (\cos\theta a_{V_1^{\prime}}^{\dagger}+\sin\theta
a_{V_3^{\prime}}^{\dagger})(\cos\theta
a_{H_2^{\prime}}^{\dagger}+\sin\theta a_{H_4^{\prime}}^{\dagger})
]\nonumber\\
&&\times [(\sin\theta a_{H_1^{\prime}}^{\dagger}-\cos\theta
a_{H_3^{\prime}}^{\dagger})(\sin\theta
a_{V_2^{\prime}}^{\dagger}-\cos\theta
a_{V_4^{\prime}}^{\dagger})\nonumber\\
&& +(\sin\theta a_{V_1^{\prime}}^{\dagger}-\cos\theta
a_{V_3^{\prime}}^{\dagger})(\sin\theta
a_{H_2^{\prime}}^{\dagger}-\cos\theta a_{H_4^{\prime}}^{\dagger})
]|0\rangle\,. \label{3}
\end{eqnarray}
If no photons are detected in the output modes $H_3^{\prime}$,
$V_3^{\prime}$, $H_4^{\prime}$ and $V_4^{\prime}$, this quantum
state is projected into
\begin{eqnarray}
\Psi_2=\frac{1}{2\sqrt{3}}(a_{H_1^{\prime}}^{\dagger}a_{V_2^{\prime}}^{\dagger}+
a_{V_1^{\prime}}^{\dagger}a_{H_2^{\prime}}^{\dagger} )^2|0\rangle
\,. \label{4}
\end{eqnarray}
The related probability of success $3\sin^4\theta\cos^4\theta$
will approach its maximum value $3/16$, if symmetric beam
splitters are used. The quantum state $\Psi_2$ can also be written
in the form
\begin{eqnarray}
\Psi_3=\frac{1}{\sqrt{3}}(|2H_1^{\prime}\rangle|2V_2^{\prime}\rangle
+|H_1^{\prime}\rangle|V_1^{\prime}\rangle|H_2^{\prime}\rangle|V_2^{\prime}\rangle
+|2V_1^{\prime}\rangle|2H_2^{\prime}\rangle)\,, \label{5}
\end{eqnarray}
which is equivalent to the quantum state (\ref{1}). Now we will
demonstrate that the quantum state (\ref{1}) can be used to
generate a telecloning state and to implement quantum telecloning
protocol. In \cite{mur} a telecloning protocol was proposed in
order to copy a quantum state from one input to $M$ outputs. In
the following we will show, how $1\rightarrow2$ optimal quantum
cloning can be employed by the use of quantum state (\ref{1}). The
experimental setup is shown schematically in Fig.2. Let mode $H_1$
and $V_1$, $H_2$ and $V_2$ pass through two symmetric beam
splitters $BS$. The state of the system evolves into
\begin{eqnarray}
\Psi_4&=&\frac{1}{8\sqrt{3}}[(a_{H_1^{\prime}}^{\dagger}+a_{H_2^{\prime}}^{\dagger})^2
(a_{V_3^{\prime}}^{\dagger}+a_{V_4^{\prime}}^{\dagger})^2+
(a_{V_1^{\prime}}^{\dagger}+a_{V_2^{\prime}}^{\dagger})^2
(a_{H_3^{\prime}}^{\dagger}+a_{H_4^{\prime}}^{\dagger})^2\nonumber\\
&&+2(a_{H_1^{\prime}}^{\dagger}+
a_{H_2^{\prime}}^{\dagger})(a_{V_1^{\prime}}^{\dagger}+a_{V_2^{\prime}}^{\dagger})(a_{H_3^{\prime}}^{\dagger}+a_{H_4^{\prime}}^{\dagger})
(a_{V_3^{\prime}}^{\dagger}+a_{V_4^{\prime}}^{\dagger})
]|0\rangle\,. \label{6}
\end{eqnarray}
If we consider only those terms, which correspond to a 4-photon
coincidence (one photon in each of the beams), the state of the
system is projected into
\begin{eqnarray}
\Psi_5&=&\frac{1}{\sqrt{3}}[
|H_1^{\prime}\rangle|H_2^{\prime}\rangle|V_3^{\prime}\rangle|V_4^{\prime}\rangle
+|V_1^{\prime}\rangle|V_2^{\prime}\rangle|H_3^{\prime}\rangle|H_4^{\prime}\rangle
\nonumber\\
&& +\frac{1}{2}(
|H_1^{\prime}\rangle|V_2^{\prime}\rangle+|V_1^{\prime}\rangle|H_2^{\prime}\rangle)
(|H_3^{\prime}\rangle|V_4^{\prime}\rangle+|V_3^{\prime}\rangle|H_4^{\prime}\rangle)]\,.
\label{7}
\end{eqnarray}
This is exactly a telecloning quantum state, which can be used to
perform $1\rightarrow2$ optimal quantum cloning \cite{mur}.\\
If $N$ pairs of polarization entangled photon states of the form
$\frac{1}{\sqrt{2}}(|H_{i}\rangle|V_{i+1}\rangle+|V_i\rangle|H_{i+1}\rangle)$
($i=1,3,\cdots,2N-1$) are prepared, we let the modes $H_i, V_i,
H_{i+1}$ and $V_{i+1}$ ($i=1,3,\cdots,2N-1$) pass through four
symmetric $N$-port beam splitters \cite{zzze}. At the output mode
of each $N$-port symmetric beam splitter $N-1$ detectors are used
to detect photons from the second output mode to the $N$th output
mode. If all detectors do not detect a photon, the state of the
system is projected to the entangled $2N$-photon state
\begin{eqnarray}
\Psi_6&=&\frac{1}{\sqrt{N}}(|NH\rangle_1|NV\rangle_2+\cdots+|mH\rangle_1|(N-m)V\rangle_1|(N-m)H\rangle_2|mV\rangle_2\nonumber\\
&&+\cdots+|NV\rangle_1|NH\rangle_2)\,.
\end{eqnarray}
This quantum state corresponds to the quantum state of two
maximally entangled spin-$\frac{N}{2}$ particles. In the
theoretical work by Drummond \cite{pdd} such multi-photon states
were used
to violate the Bell inequality.\\
In conclusion, a scheme is presented to produce an entangled
four-photon state from two pairs of entangled two-photon states.
Such entangled four-photon states are equivalent to the state of
two maximally entangled spin-1 particles. We further show that
such entangled four-photon states can also be used to generate a
telecloning state. This state has been used to implement
$1\rightarrow2$ optimal quantum cloning. This scheme can be
generated in order to prepare an entangled $2N$-photon state from
$N$ pairs of entangled two-photon states. Such multi-photon states
may play a crucial role in fundamental tests of quantum mechanics
versus local realism and in many quantum information and quantum
computation schemes.

\begin{flushleft}

{\Large \bf Figure Captions}

\vspace{\baselineskip}

{\bf Figure 1.} The schematic of the entangled four-photon state
generation from a pair of polarization entangled two-photon states
is shown. $BS_i$ denotes the symmetric beam splitters and $D_i$
are photon number detectors.

{\bf Figure 2.} The schematic of the telecloning  state generation
from an entangled four-photon state is shown. $BS_i$ and $D_i$
denote the symmetric beam splitters and photon number detectors.
\end{flushleft}

\end{document}